\documentclass[preprint2]{aastex}

\slugcomment{Not to appear in Nonlearned J., 45.}

\shorttitle{Collapsed Cores in Globular Clusters}
\shortauthors{Djorgovski et al.}

\begin{document}


\title{Bose-Einstein condensation of photons in the matter-domonated universe}

\author{Ze  Cheng}
\affil{School of Physics, Huazhong University of Science and
   Technology, Wuhan 430074, People's Republic of China}
\email{zcheng@mail.hust.edu.cn}

\begin{abstract}
In 1914, Planck introduced the concept of a white body. In nature,
no true white bodies are known. We assume that the universe after
last-scattering is an ideal white body that contains a tremendously
large number of thermal photons and is at an extremely high
temperature. Bose-Einstein condensation of photons in an ideal white
body is investigated within the framework of quantum statistical
mechanism. The computation shows that the transition temperature
$T_c$ is a monotonically increasing function of the number density
$n$ of photons. At finite temperature, we find that the condensate
fraction $N_0(T)/N$ decreases continuously from unity to zero as the
temperature increases from zero to the transition temperature $T_c$.
Further, we study the radiation properties of an ideal white body.
It is found that in the condensation region of $T<T_c$, the spectral
intensity $I(\omega,T)$ of white body radiation is identical with
Planck's law for blackbody radiation.
\end{abstract}



\section{Introduction}

Nowadays it is recognized that the Bose-Einstein condensation (BEC)
is a common quantum property of the many-particle systems in which
the number of particles is conserved. If particles are bosons, the
interaction between particles must be repulsive, whereas if
particles are fermions, the interaction must be attractive. The
generalized BEC is the macroscopic accumulation of Bose particles in
the energetic ground state below a critical temperature. In 1995,
the three research groups in the U. S. A. observed the BEC of
ultracold Bose atomic gases in a trapping potential
\citep{and95,dav95,bra95}. In 1938, Kapitza discovered the
superfluid phase of liquid $^4$He below 2.17 K, which can be viewed
as a BEC among strongly interacting $^4$He atoms. In 1911, Onnes
discovered the superconducting state of metal mercury below 4 K. In
1957, Bardeen, Cooper, and Schrieffer provided a successful
microscopic description of superconductors in terms of Cooper pairs
\citep{coo56,bar57}. The superconducting state can be regarded as a
manifestation of the BEC of Cooper pairs. The BEC has been observed
also in several systems of solid-state quasiparticles, which include
excitons \citep{but02,eis04}, exciton-polaritons \citep{den10}, and
magnons \citep{nik00,dem06}.

However, the most omnipresent blackbody radiation does not show this
phase transition. The photons in blackbody radiation have a
vanishing chemical potential, so that the number of photons is not
conserved when the temperature of the blackbody is varied; at low
temperatures, photons are absorbed by the walls of the blackbody
instead of occupying the zero-momentum state. Theoretical works have
considered thermalization processes that conserve photon number,
involving Compton scattering with a gas of thermal electrons
\citep{zel69} or photon-photon scattering in a nonlinear resonator
configuration \citep{chi99,chi00,bol01}. Number-conserving
thermalization was experimentally observed \citep{kla10a} for a
two-dimensional photon gas in a dye-filled optical microcavity,
which acts as a `white-wall' box. In the presence of thermalization
processes that conserve photon number, Weitz and colleagues have
observed the BEC of two-dimensional photons in a dye-filled optical
microcavity \citep{kla10b,kla12}. Kirton and Keeling have
established a nonequilibrium model of photonic BEC in the dye-filled
microcavity \citep{kir13}.

Now, let us turn to a gas of three-dimensional photons in the
universe. The evolution of the universe according to the standard
hot big-bang model \citep{tur99} is sketched as follows: (1)
Radiation-dominated phase. At the age of the universe earlier than
about 10000 years, when the temperature of the universe exceeded
$k_B T\gtrsim 3$ eV, the energy density in radiation and
relativistic particles exceeded that in matter. At the earliest
times, the energy in the universe consists of radiation and seas of
relativistic particle-antiparticle pairs. At the time of 10$^{-11}$
second, when $k_B T\sim 300$ GeV, the sea of relativistic particles
includes six species of quarks and antiquarks, six types of leptons
and antileptons, and twelve gauge bosons. When $k_B T\lesssim 2mc^2$
where $m$ is the mass of a particle species, those particles and
their antiparticles annihilate into at least two photons. When the
universe was seconds old and the temperature was around 1 MeV,
big-bang nucleosynthesis led to the production of the light elements
D, $^3$He, $^4$He, and $^7$Li. When $k_B T\ll 1$ MeV, the last of
the particle-antiparticle pairs, the electrons and positrons,
annihilated. (2) Matter-dominated phase. When the temperature
reached around $k_B T\sim 3$ eV, at a time of around 10000 years the
energy density in matter began to exceed that in radiation. Shortly
after matter domination begins, photons in the universe undergo
their last-scattering off free electrons. Last scattering is
precipitated by the recombination of electrons and ions (mainly free
protons), which occurs at a temperature of $k_B T\sim 0.3$ eV
because neutral atoms are energetically favored. Before
last-scattering, matter and radiation are tightly coupled; after
last-scattering, matter and radiation are essentially decoupled.
Under the surface of last scattering, the number of photons in the
universe is conserved.

In 1914, Planck introduced the concept of a white body
\citep{pla14}. A white body is one for which all incident radiation
is reflected uniformly in all directions, an idealization exactly
opposite to that of the blackbody. In nature, no true white bodies
are known. We assume that the universe after last-scattering is an
ideal white body that contains a tremendously large number of
thermal photons and is at an extremely high temperature. As shown in
Fig. 1, an ideal white body can be regarded as a rectangular cavity
whose walls are high-reflecting planar optical mirrors and are kept
at a constant temperature $T$. This white body contains only thermal
radiation in its interior. Because the cavity walls absorb no
electromagnetic radiation, the number of photons is conserved in the
white body. In the present paper, we shall study the BEC of photons
in an ideal white body. The BEC of photons in an ideal white body
takes place in the momentum space. The planar mirrors provide a
chemical potential for a photon, making the system formally
equivalent to a three-dimensional gas of number-conserving, massless
bosons. The advantages of an ideal white body are as follows: (1)
the loss rate is zero and (2) the nonlinearity does not exist.

\begin{figure}[t]
  \includegraphics[width=7.9cm]{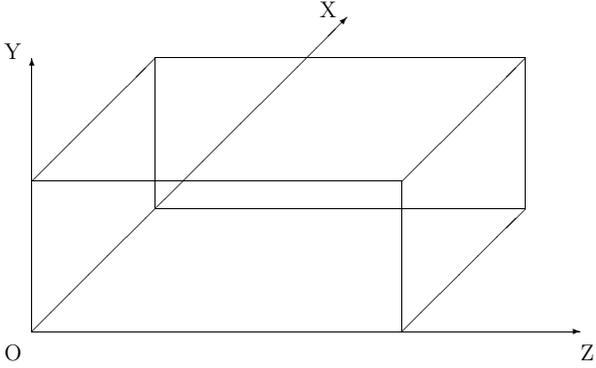}
  \caption{An ideal white body: a rectangular cavity
enclosed by perfectly conducting walls and kept at a constant
temperature.}
\end{figure}

The important properties of such a BEC will be expounded in the
following. At first, we investigate the BEC of noninteracting
photons in an ideal white body. It is found that such a BEC is a
second phase transition. The expresson of transition temperature
$T_c$ is obtained, which is dependent on the number density $n$ of
photons. At finite temperature, we find that the condensate fraction
$N_0(T)/N$ decreases continuously from one to zero as the
temperature increases from zero to the transition temperature $T_c$.
We find that the vacuum we currently observe is the condensate or
superfluid of photons. Further, we study the radiation properties of
an ideal white body. It is found that in the condensation region of
$T<T_c$, the spectral intensity $I(\omega,T)$ of white body
radiation is identical with Planck's law for blackbody radiation.
Our investigation into the BEC of three-dimensional photons deepens
our understanding of the development of the universe as an ideal
white body. Since the BEC of three-dimensional photons was never
explored previously, we point out here that there are new features,
which we believe are worthy of exploration. The predicted properties
of the BEC of photons in an ideal white body are highly relevant to
the present universe.

\section{Formation of photonic Bose-Einstein condensates}

Because the number of photons is conserved in a white body, the
photon system possesses a nonvanishing chemical potential $\mu$. Now
we need to quantize the electromagnetic field. In terms of the
creation and annihilation operators $a^{\dagger}_{{\bf k}\sigma}$
and $a_{{\bf k}\sigma}$, respectively, of circularly polarized
photons with wave vector $\bf k$ and helicity $\sigma=\pm 1$, the
grand canonical Hamiltonian ${\cal H}_{em}$ of the photon system is
given by
\begin{equation}
{\cal H}_{em}=\sum_{{\bf k}\sigma}(\hbar\omega_{\bf
k}-\mu)a^{\dagger}_{{\bf k}\sigma} a_{{\bf k}\sigma}\,.
\end{equation}
where the zero-point energy terms are dropped. Here $\hbar$ is
Planck's constant reduced, $\omega_{\bf k}=c|{\bf k}|$ is the
angular frequency of a photon, and $c$ is the propagation velocity
of light in the vacuum. Equation (1) represents the grand canonical
Hamiltonian of the system of noninteracting photons.

In the representation with a set of quantum numbers $i\equiv\{{\bf
k}, \sigma\}$, an energy level of a three-dimensional photon is
denoted by the quantum number $i$. In the grand canonical ensemble,
the system under study consists of $N$ noninteracting photons, which
are distributed over various quantum states $i$ and have a chemical
potential $\mu$. Based on the first principles of statistical
mechanics, one knows that the average number $\langle N_i\rangle$ of
photons in the $i$th state of energy $E_i$ obeys the Bose-Einstein
distribution
\begin{equation}
\langle N_i\rangle={1\over e^{\beta(E_i-\mu)}-1}\,,
\end{equation}
where $\beta$ is related to the temperature $T$ by $\beta=1/k_BT$
and $k_B$ is the Boltzmann constant. The chemical potential $\mu$ is
determined by the constraint that the total number of photons in the
system is $N$:
\begin{equation}
\sum_i \langle N_i\rangle=N\,.
\end{equation}
The phenomenon of BEC for noninteracting photons is fully described
by Eqs. (2) and (3). The nontrivial aspect is the determination of
the chemical potential as a function of $N$ and $T$. Once $\mu$ is
known, all thermodynamic quantities like total energy, specific
heat, and pressure follow directly from sums over the energy levels
involving the occupation numbers in Eq. (2).

To this end, Eq. (2) can be rewritten as follows:
\begin{equation}
\langle N_i\rangle={ze^{-\beta E_i}\over 1- ze^{-\beta E_i}}\,,
\end{equation}
where the fugacity $z$ can be expressed as $z=\exp(\beta\mu)$ and
$E_i=E_{\bf k}=\hbar\omega_{\bf k}$. The ground state of the system
is the state with ${\bf k}={\bf 0}$, and the energy of the ground
state is zero. The BEC means that at low temperatures, a macroscopic
number of photons occupies the ground state. From Eq. (4), the
average number of photons in the ground state is $N_0=2z/(1- z)$.
The ground-state population diverges as $z\rightarrow 1$. When we
split off the the ground-state population in Eq. (3), Eq. (3)
becomes
\begin{equation}
{2z\over 1-z}+\sum_{i\neq 0}\langle N_i\rangle=N\,.
\end{equation}

At this point, we should note that the sum over $i$ in Eq. (5)
includes an integral over ${\bf k}$ and a sum over $\sigma$. On
putting Eq. (4) into Eq. (5), Eq. (5) is transformed into the
following form:
\begin{equation}
{2z\over 1-z}+ {2V\over\pi^2(\beta\hbar c)^3}g_3(z)=N\,,
\end{equation}
where $g_{3}(z)$ is a special case of the Bose functions $g_n(z)$.
It is obvious that for real values of $z$ between 0 and 1,
$g_{3}(z)$ is a bounded, positive, monotonically increasing function
of $z$. At $z=0$, $g_{3}(0)=0$, and at $z=1$, $g_{3}(1)=1.202$. To
satisfy Eq. (6), it is necessary that $0\leq z\leq 1$. The fugacity
$z$ can be determined numerically from Eq. (6). $z$ is a function of
temperature $T$, volume $V$, and photon number $N$. Once $z$ is
known, the average number of photons in the ground state can be
obtained from the relation $N_0=2z/(1-z)$.

Let us rewrite Eq. (6) in the form,
\begin{equation}
N_0+{2V\over\pi^2(\beta\hbar c)^3}g_3(z)=N\,.
\end{equation}
The critical temperature $T_c$ can now be found by setting $N_0=0$
and $z=1$ in Eq. (7). This results in the following expression for
the critical temperature,
\begin{equation}
T_c={\hbar c\over k_B}\left[n\pi^2\over 2g_{3}(1)\right]^{1/3}\,,
\end{equation}
where $n=N/V$ is the number density of photons. In the limit as
$V\rightarrow\infty$, we obtain the solution of Eq. (6),
\begin{equation}
z=\left\{ \begin{array}{ll}
1\,,\; T\leq T_c\,,\\
         \\
\mbox{the root of}\;g_{3}(z)=(T_c/T)^3g_{3}(1)\,,\; T>T_c\,.
\end{array} \right.
\end{equation}
By virtue of Eq. (9), from Eq. (7) we find that the condensate
fraction of photons in the ground state is given by
\begin{equation}
{N_0\over N}=\left\{ \begin{array}{ll} 1-\left(T\over
T_c\right)^{3}\,,
\; T\leq T_c\,,\\
         \\
0\,,\; T>T_c\,.
\end{array} \right.
\end{equation}
This means that a macroscopic number of photons occupies the ground
state. This phenomenon is known as the Bose-Einstein condensation.
The photons in the ground state form a condensate, which is
superfluid.

\begin{figure}[t]
  \includegraphics[width=7.9cm]{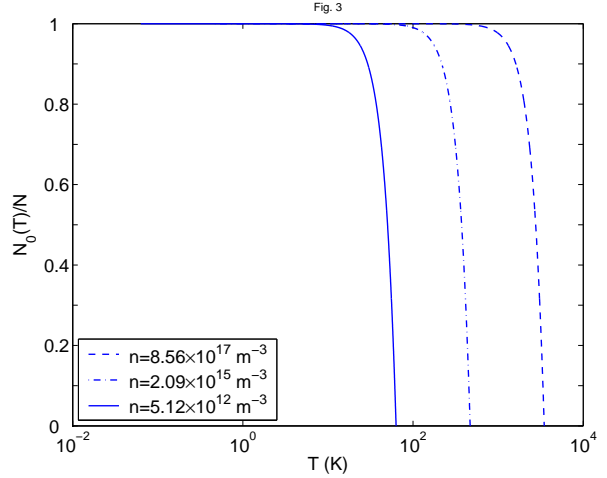}
  \caption{In an ideal white body, variation of the condensate
fraction $N_0/N$ with the temperature $T$ and the number density
$n$.}
\end{figure}

From the above description, we can see that the BEC of
three-dimensional photons in a white body takes place in the
momentum space. As we have known, last-scattering deceased at a
temperature of $k_B T\sim 0.3$ eV. The critical temperature $T_c$ of
photonic BEC should meet the condition $k_B T_c\lesssim 0.3$ eV. In
other words, this condition is $T_c\lesssim 3481.35$ K. From Eq.
(8), $T_{c}$ is directly proportional to $n^{1/3}$. At $n=5.12\times
10^{12}$, $2.09\times 10^{15}$, $8.56\times 10^{17}$ m$^{-3}$,
$T_{c}=63.20$, $469.03$, $3480.95$ K, respectively. According to Eq.
(10), the variation with $T$ and $n$ of condensate fraction $N_0/N$
is shown in Fig. 2. For a fixed $n$, the condensate fraction
decreases continuously from unity to zero as the temperature
increases from zero to the critical temperature $T_c$. For a fixed
$T$, the condensate fraction is a monotonically increasing function
of $n$.

\section{Radiation properties of an ideal white body}

The universe before last-scattering was filled with a large number
of thermal free electrons. The electromagnetic field within the
universe before last-scattering is in thermodynamical equilibrium
but has no conserved photon number \citep{tur99}. Such equilibrium
is established via the continual scattering of photons by thermal
free electrons. The electromagnetic field in thermal equilibrium is
called thermal radiation and characterized by a definite temperature
$T$. The universe after last-scattering can keep the photon number
conserving and thus becomes a white body. The thermal radiation in a
white body is also called white-body radiation. The photons in a
white body are in a thermal radiation state, which is called a
normal state. In order to characterize the thermal radiation state,
we need to conceive a grand canonical ensemble of photons. Some
identical systems of the ensemble may be in an eigenstate of the
Hamiltonian ${\cal H}_{em}$ given by Eq. (1), while the distribution
of the ensemble over the eigenstates is described by the density
operator of the thermal radiation state
\begin{equation}
\rho={\exp(-{\cal H}_{em}/k_B T)\over \mbox{Tr}\,\exp(-{\cal
H}_{em}/k_B T)}\,.
\end{equation}
The basis states used in the trace are the eigenstates of the
Hamiltonian ${\cal H}_{em}$. The main thermodynamic quantity in
white-body radiation is the total energy $E$ or the energy density
$u=E/V$, which is the ensemble average of the corresponding
microscopic quantity,
\begin{equation}
E=\sum_{{\bf k}\sigma}\hbar\omega_{\bf k}\langle N_{{\bf
k}\sigma}\rangle\,.
\end{equation}
Here we have utilized the average notation $\langle N_{{\bf
k}\sigma}\rangle =\mbox{Tr}(\rho N_{{\bf k}\sigma})$.

It is easily found that the ensemble average of the number operator
of photons in a mode ${\bf k}\sigma$ satisfies the well-known
Bose-Einstein distribution given by Eq. (2). In the usual way
altering the summation to an integration, we obtain
\begin{equation}
E=V\int^{\infty}_0\rho(\omega,T)d\omega\,,
\end{equation}
\begin{equation}
\rho(\omega,T)={\hbar\over\pi^2c^3}{\omega^3\over z^{-1}
e^{\hbar\omega/k_B T}-1}\,,
\end{equation}
where $z=\exp(\mu/k_B T)$ is the fugacity. Equation (14) gives the
spectral energy density of white body radiation.

The spectral energy density $\rho(\omega,T)$ of white body radiation
is not an observable. An observable of white body radiation is the
spectral intensity $I(\omega,T)$ of white body radiation.
$I(\omega,T)$ is defined as the power per unit surface area per unit
solid angle per unit frequency emitted at a frequency $\omega$ by a
white body. $I(\omega,T)$ is related to $\rho(\omega,T)$ through the
relation:
\begin{equation}
I(\omega,T)={c\over 4\pi}\rho(\omega,T)={\hbar\over
4\pi^3c^2}{\omega^3\over z^{-1} e^{\hbar\omega/k_B T}-1}\,.
\end{equation}
In the condensation region of $T<T_c$, $z=1$ and so $I(\omega,T)$ is
identical with Planck's law for blackbody radiation. In the
non-condensation region of $T>T_c$, $z<1$ and so $I(\omega,T)$
becomes smaller than Planck's law for blackbody radiation at the
same frequency and temperature. At $T=T_c$, $I(\omega,T)$ is
continuous.

On putting Eq. (14) into Eq. (13), we obtain the following result:
\begin{equation}
E(T)={4\sigma\over c}VT^4{g_4(z)\over g_4(1)}\,,
\end{equation}
where $\sigma=\pi^2k_B^4/60\hbar^3c^2$ is called the
Stefan-Boltzmann constant and $g_4(1)=\pi^4/90$. The energy density
$u(T)$ of white body radiation is not an observable. Another
observable of white body radiation is the total intensity $R(T)$ of
white body radiation. $R(T)$ is defined as the total energy radiated
per unit surface area of a white body per unit time. $R(T)$ is
related to $u(T)$ through the relation:
\begin{equation}
R(T)={c\over 4}u(T)=\sigma T^4{g_4(z)\over g_4(1)}\,.
\end{equation}
In the condensation region of $T<T_c$, $z=1$ and so $R(T)$ is
identical with the Stefan-Boltzmann law for blackbody radiation. In
the non-condensation region of $T>T_c$, $z<1$ and so $R(T)$ becomes
smaller than the Stefan-Boltzmann law for blackbody radiation. At
$T=T_c$, $R(T)$ is continuous. It is interesting and important to
note that all of the energy density arises from the nonzero mementum
photons. That is, zero-momentum photons, however numerous,
contribute nothing to $u(T)$.

\begin{figure}[t]
  \includegraphics[width=7.9cm]{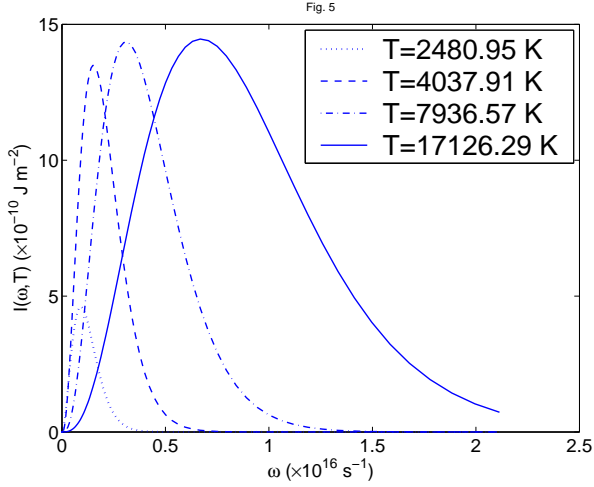}
  \caption{In an ideal white body, at a fixed density
$n=8.56\times 10^{17}$ m$^{-3}$, variation of the spectral intensity
$I(\omega,T)$ with the frequency $\omega$ and the temperature $T$.}
\end{figure}

\begin{figure}[t]
  \includegraphics[width=7.9cm]{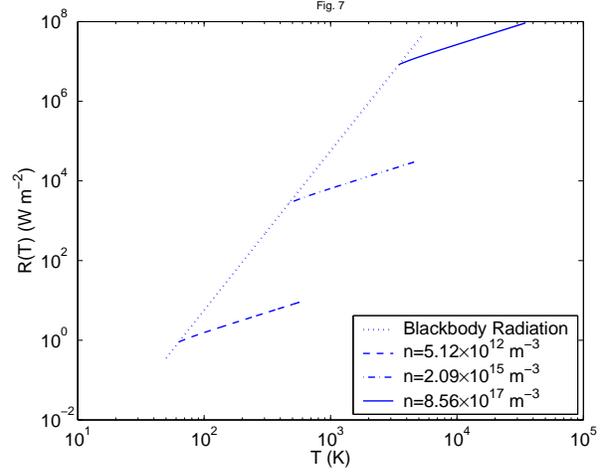}
  \caption{In an ideal white body, variation of the total intensity $R(T)$
  with the temperature $T$ and the density $n$. The dotted line
represents variation of the total intensity $R(T)$ of blackbody
radiation with the temperature $T$}
\end{figure}

For convenience, in Eq. (15) we set the number density of photons
$n=8.56\times 10^{17}$ m$^{-3}$. The transition temperature $T_c$
depends on the number density $n$. $T_{c}=3480.95$ K at
$n=8.56\times 10^{17}$ m$^{-3}$. Further, the fugacity $z$ depends
on the temperature $T$. We find that at $T=2480.95$, $4037.91$,
$7936.57$, $17126.29$ K, $z=1.0, 0.7, 0.1, 0.01$, respectively. The
spectral intensity of white body radiation at density $n=8.56\times
10^{17}$ m$^{-3}$ is plotted in Fig. 3 as a function of its
frequency, where dotted, dashed, and dashed-dotted, and solid lines
correspond to $T=2480.95$, $4037.91$, $7936.57$, $17126.29$ K,
respectively. There are three features: (i) in the condensation
region of $T<T_c$, $I(\omega,T)$ is identical with Planck's law for
blackbody radiation; (ii) in the non-condensation region of $T>T_c$,
$I(\omega,T)$ becomes larger and larger than Planck's law for
blackbody radiation as temperature $T$ ascends; (iii) the peak
frequency of blackbody radiation at $T=2480.95$ K is
$\omega_m=9.2\times 10^{14}$ s$^{-1}$ while the peak frequency of
white body radiation at $T>T_c$ becomes larger and larger as
temperature $T$ ascends. Next let us turn to Eq. (17). The total
intensity of white body radiation is plotted in Fig. 4 as a function
of its temperature, where the dashed, dashed-dotted, and solid lines
correspond to $n=5.12\times 10^{12}$, $2.09\times 10^{15}$,
$8.56\times 10^{17}$ m$^{-3}$ , respectively. The dotted line
represents variation of the total intensity of blackbody radiation
with the temperature. In the non-condensation region of $T>T_c$,
$R(T)$ becomes smaller than the Stefan-Boltzmann law for blackbody
radiation. At a fixed density $R(T)$ is a monotonically increasing
function of $T$ and at a fixed temperature $R(T)$ is also a
monotonically increasing function of $n$.

\section{Discussion}

In investigating the BEC of three-dimensional noninteracting
photons, we assume that the universe after last-scattering is an
ideal white body that contains a tremendously large number of
thermal photons ($n\lesssim 8.56\times 10^{17}$ m$^{-3}$) and is at
an extremely high temperature ($T\lesssim 3481.35$ K). This BEC
theory is highly relevant to the present universe. The vacuum state
of the universe before last-scattering is a true vacuum state of
photons, in which $a_{{\bf k}\sigma}|0\rangle=0$ for all ${\bf k}$.
As the temperature of the universe descends, the universe after
last-scattering undergoes a second-order phase transition into a BEC
state. Consequently, the white-body radiation in the universe
becomes the blackbody radiation. The vacuum state of the present
universe is a condensate or superfluid of photons. This vacuum state
is defined as follows:
\begin{equation}
|N_0\rangle={1\over \sqrt{N_0!}} (a^{\dagger}_{{\bf
0}\pm})^{N_0}|0\rangle\,,
\end{equation}
where $N_0$ is given by Eq. (10). This vacuum state satisfies the
property: $a_{{\bf k}\sigma}|N_0\rangle=0$ for all ${\bf k}\ne {\bf
0}$. The observational findings in astronomy \citep{rie98,per99} can
be correctly interpreted in terms of a BEC theory of noninteracting
thermal photons. Such a theory can account for fluctuations of the
vacuum and for the cosmic microwave background radiation
\citep{pen65}.

The photons in the state $|N_0\rangle$ are virtual photons, which
can not be detected by any means. Because of fluctuations of the
vacuum, virtual photons becomes real photons with extremely short
lifetime. Therefore, the vacuum state possesses the vacuum energy
$E_0$, as given by
\begin{equation}
E_0={\sum_{{\bf k}\sigma}}'{1\over 2}\hbar\omega_{\bf k}\,,
\end{equation}
where the prime on the summation symbol means that $|{\bf k}|\leq
k_m$. $k_m$ is the maximum wave number of virtual photons and must
satisfy the relation $k_m\ll m_e c/\hbar$, where $m_e$ is the rest
mass of electron. In the usual way altering the summation to an
integration, we obtain $E_0=Vu_0$, where $u_0$ is the energy density
of the vacuum and is given by $u_0={\hbar ck_m^4/8\pi^2}$. Although
$u_0$ is very small, the volume $V$ of the universe is enormous, so
that $E_0$ is tremendously large. The vacuum energy $E_0$ is called
the dark energy, because it can not be detected by any means. The
dark energy perhaps drives the accelerating expansion of the
universe \citep{rie98,per99}. In 1998, two research teams studied
several dozen, distant type Ia supernovae and discovered that the
universe is expanding at an ever-accelerating rate
\citep{rie98,per99}. The discovery of cosmic acceleration is
arguably one of the most important developments in modern cosmology.
The physical origin of cosmic acceleration remains a deep mystery. A
probable explanation of cosmic acceleration is that 75\% of the
energy density of the universe exists in a new form with large
negative pressure, known as dark energy \citep{tur99,hut99}. Dark
energy is dark because its action is unknown. Two proposed forms for
dark energy are the cosmological constant, a constant energy density
filling space homogeneously \citep{pee03,wei89}, scalar field models
\citep{cam11} which comprehend, e.g., quintessence, phantom,
K-essence, and tachyon fields.

Within the framework of quantum statistical mechanism, we
investigate the BEC of three-dimensional photons in an ideal white
body, which acts as a model of the universe after last-scattering.
During the radiation-dominated stage of the development of the
universe, the three-generation charged leptons are a unitary Dirac
field. In the meanwhile, the electromagnetic field and the Dirac
field are in thermal equilibrium with each other. Under the surface
of last scattering, the number of photons in the universe is
conserved. The electromagnetic field in thermal equilibrium and with
a conserved photon number is called a white-body radiation. After
the temperature of the universe becomes smaller than 3481.35 K, the
universe undergoes a second-order phase transition into a BEC state.
Consequently, the white-body radiation in the universe becomes the
blackbody radiation that we observe in the present universe. The
vacuum that we currently observe is a condensate consisting of
zero-momentum photons. This phase transition occurs in the
matter-dominated stage of the development of the universe. The
transition temperature of $3480.95$ K is much higher than any
temperature we normally encounter in the present universe.

In summary, we have proposed a BEC theory of three-dimensional
photons in the universe after last-scattering, which can be regarded
as an ideal white body. At zero temperature, we find that all
photons in the universe condense into the zero-momentum state. The
computation shows that the transition temperature $T_c$ is about
$3480.95$ K. The photon system undergoes a second-order phase
transition from the normal state to the BEC state. In the meanwhile,
the white-body radiation in the universe becomes the blackbody
radiation that we observe in the present universe. The predicted
properties of the BEC of ideal thermal photons are highly relevant
to the present universe.

\acknowledgments

This work was supported by the National Natural Science Foundation
of China under Grants No. 10174024 and No. 10474025.

%% in parentheses, after the keyword, but they will not be verified.

%% Three table samples follow, two marked up in the deluxetable environment,
%% one marked up as a LaTeX table.

%% This table also includes a table comment indicating that the full
%% version will be available in machine-readable format in the electronic

\clearpage

\end{document}